\documentstyle[preprint,aps]{revtex}

\newcommand{\beq}{\begin{equation}}
\newcommand{\eneq}{\end{equation}}

\newcommand{\vecr}{\vec{r}}

\begin{document}

\begin{titlepage}

\baselineskip=16pt

\title{ A dual 2D model for the Quantum Hall Fluid
\footnote{Work supported in part by MURST and by EEC 
contract no. SCI-CT92-0789}}
\author{$G. Cristofano^{*,\dagger}$,$D. Giuliano^{*,\dagger,\%}$ and
$G. Maiella^{*,\dagger}$}
\address{$~^*$ INFN Sezione di Napoli\\
$~^\dagger$ Dipartimento di Scienze Fisiche, Universita' di Napoli
I-80125 Napoli Italy\\
$~^\%$ Department of Physics, Stanford University, Stanford, CA 94305}
\date{\today}  
\maketitle
\begin{abstract}
We present a dual 2D statistical model of a Quantum Hall Fluid
which depends on a coupling constant $g$
and an angular variable $\theta$, parametrizing a surface term. We 
show 
that such a model has topologically non trivial vacua (corresponding 
to rational values of the filling), which are 
infrared stable fixed points of the renormalization group. Moreover 
its partition function has a dual infinite discrete symmetry, 
$SL(2,Z)$, which reproduces the phenomenological laws of corresponding 
states. Such a symmetry reproduces the universality of the phase 
diagram of the Hall Fluid and allows for an unified description of 
its fixed points in terms of a 2D Conformal Field Theory with central 
charge $c=1$. 
\end{abstract}

\end{titlepage}

\setcounter{page}{1}
\newpage

Our understanding of the 
conduction  properties of a Quantum Hall Fluid at the plateaux
has greatly improved recently by means of the 
two-dimensional Conformal Field Theory (2D CFT) \cite{1}. In 
particular a theoretical framework has been proposed,
 which basically 
substantiates Laughlin physical idea of associating a magnetic flux to 
a charge identifying the anyons
 as basic excitations \cite{2}. Furthermore both the Integer Quantum Hall 
Effect and the Fractional one 
now appear strictly related to each other, sharing similar many-body 
properties. 
Then we 
can ask the question of the form of the phase diagram and its critical
points \cite{3,4}.  
 In this context there has been a theoretical as well as a phenomenological 
attempt \cite{3,5} to uncover the physical properties of such a diagram: its 
dependence on two parameters (and their physical meaning), 
its nested character and the universality shared
by the fixed points (attractive or repulsive) describing the Hall 
fluid at different fillings.

We are able to build a 2D statistical model, which, besides the usual 
g-dependance of the kynetic term and a surface term, parametrized by 
the angle $\theta$,  
\cite{6,7}, contains newly generated electric and magnetic background 
terms.

We extend the Renormalization Group (RG) analysis to such 
a model sorting out its new interesting properties, as, 
for example, 
the existence of Infrared (IR) stable fixed points 
which are well described 
by the presence of both electric and magnetic backgrounds. These are 
interpreted as topologically non-trivial vacua of the field theory in 
the continuum which takes the form of a generalized dual Coulomb Gas 
with background, reproducing, for rational values of $\theta$, the 
phenomenologically observed properties of a Hall fluid at the plateaux 
at corresponding fillings.

Also we observe the existence of an infinite discrete 
symmetry $SL(2,Z)$ (generalizing the well known Kramers-Wannier duality)
which acts as in ref. \cite{7}, mapping those 
non trivial fixed points one into another. This suggests a unified 
picture of the IR fixed points in terms of a 2D Conformal Field 
Theory with central charge $c=1$.

We stress that there is a simple connection between the duality 
transformations of the model and the 
phenomenological "laws of corresponding states" advocated by the 
authors of ref. \cite{5}.

Let us first remind how Cardy and Rabinovici build their 2D model 
starting from a $U(1)$ gauge theory with both electric and magnetic 
matter coupled by a surface $\theta$ term \cite{6}. Then we will search for 
a non trivial extension of such a model in which a fixed background 
is generated. 

The explicit action is 
given by:
\beq
S [ A, S,  n ] \equiv \frac{1}{2g}\int d^2r(\partial_\beta A - 
S_\beta)^2 - i\int d^2r n A - i\frac{\theta}{2\pi}\int d^2r 
\epsilon_{\beta\gamma}S_\beta \partial_\gamma A ~~,
\eneq
where $A$ is the gauge potential,
$S_\beta(\vecr)$ is the ``magnetic frustration field", defined by 
the constraint:
\beq
\epsilon_{\beta\gamma}\partial_\beta S_\gamma (\vecr) - m(\vecr) = 0,
\eneq

while $n(\vecr)$ and $m(\vecr)$ are defined in terms of the electric and 
magnetic charge density; $\epsilon_{\beta\gamma}$ is the 
antisymmetric tensor in 2D.
 In particular, the magnetic charge density is 
$Q_m = m/ \sqrt{g}$ while the electric one is obtained by integrating 
out the field $S_\beta$ with the constraint (2) and is given by:

\[  
Q_e = \sqrt{g} \left( n +\frac{\theta}{2\pi} m\right)~.
\]

The densities ($n$, $m$) are constrained by the neutrality condition (required 
to make the system infrared stable):
\beq
\int d^2 r~ n(\vecr) = \int d^2 r ~m (\vecr) = 0 ~~.
\eneq
In the following we need two representations of our model, one as a
Coulomb gas where the role of the electric and
magnetic charges is emphasized,
 and the gauge representation, where both charges have gauge
interaction. 

The Coulomb gas representation is defined by:
\beq
e^{-S_{CG}(n, m)} = \int {\cal D}A \int \prod_{\alpha=1,2} {\cal 
D}S_\alpha \delta (\epsilon_{\beta\gamma}\partial_\beta S_\gamma - m)
e^{- S [ A, S , n]}~~.
\eneq
It is straightforward to evaluate the path integrals in eq.(4) and the 
result is given by:
\begin{eqnarray}
S_{CG} [ n, m] = \frac{g}{2}\int d^2r d^2r^{'} (n(\vec{r}) + 
\frac{\theta}{2\pi} m(\vec{r})) (n(\vec{r}^{'}) + 
\frac{\theta}{2\pi} m(\vec{r}^{'})) G(\vec{r}-\vec{r}^{'})+\\
\nonumber
\frac{1}{2g}\int d^2r d^2r^{'} m(\vec{r})  m(\vec{r}^{'}) G(\vec{r}-\vec{r}^{'})+
i \int d^2r d^2r^{'} n(\vec{r}) m(\vec{r}^{'})\varphi (\vec{r} - 
\vec{r}^{'})~~,
\end{eqnarray}
where $G(\vec{r})$ and $\varphi(\vec{r})$ are the ``longitudinal" and 
``transverse" Green-Feynman functions in 2D given by:
\beq
G(\vec{r}) = \ln\left(\frac{|\vecr|}{a}\right) ~~,~ \varphi(\vec{r}) = \arctan 
\left(\frac{y}{x}\right) ~~,  
\eneq
where $a$ is a cutoff.
The last term in eq.(5) is the (imaginary) Bohm-Aharonov
term \cite{8}. Also notice that eq.(5) defines for $ \theta = 0$ the standard 
Coulomb gas for both electric and magnetic charges( see ref.[9]).

To obtain the gauge representation we solve the constraint (2) as:
\beq
\delta (\epsilon_{\beta\gamma} \partial_\beta S_\gamma - m) = \int {\cal D}
A_D \exp\left( -i\int d^2r A_D(r)(\epsilon_{\beta\gamma}\partial_\beta
S_\gamma - m)\right) ~~.
\eneq
Then 
by evaluating explicitly the functional integral in eq.(4) 
 one gets the  action:

\beq
S [ A, A_D ] =\frac{g}{2} \int d^2 r (\partial_\beta A_D)^2 - i\int 
d^2r \epsilon_{\beta\gamma} \partial_\beta A \partial_\gamma A_D ~~. 
\eneq

It turns out that the relevant 2-points functions are:

\[
\langle A(\vec{r}) A(\vec{r}^{'}) \rangle = g G(\vec{r} - \vec{r}^{'}) ~~,
\hspace{5.2cm}(9.a)
\]

\[
\langle A_D(\vec{r}) A_D(\vec{r}^{'}) \rangle = \frac{1}{g} G(\vec{r} - \vec{r}
^{'}) ~~,  
\hspace{5cm}(9.b)
\]

\[
\langle A(\vec{r}) A_D(\vec{r}^{'}) \rangle = i \varphi(\vec{r} - \vec{r}^{'}) 
 ~~.
\hspace{5.2cm}(9.c)
\]
 
The dual form of the Green functions in eqs.(9.a), (9.b) should be noticed.
Naturally one can recover the Coulomb gas representation eq.(5) by choosing 
for the charge densities: 

\setcounter{equation}{9}
\beq
 n(\vecr) = \sum_i n_i \delta( \vec{r} -\vec{r}_i)~;~
 m(\vecr) = \sum_i m_i \delta( \vec{r} -\vec{r}_i) ~~.
\eneq

We now search for solutions where the gauge fields can be splitted in 
two parts: a `background" one and a ``fluctuating" one, as:

\beq
A(\vecr) \equiv \bar{A} (\vecr) + a(\vecr) ~~,~ \partial^2 \bar{A} (\vecr) = -ig 
(\bar{n} +\frac{\theta}{2\pi} \bar{m})\eneq

\vspace{0.5cm}

\begin{eqnarray}
A_D(\vecr) \equiv \bar{A}_D (\vecr) + a_D(\vecr) ~~,~ \partial^2 \bar{A}_D (\vecr) = 
\frac{i}{g}\bar{m}
\\
S_\beta(\vecr) \equiv \bar{S}_\beta (\vecr) + s_\beta(\vecr) ~~,~ 
\epsilon_{\beta\gamma }\partial_\beta \bar{S}_\gamma (\vecr) = \bar{m} 
~~.\nonumber 
\end{eqnarray}
The equations above  can be easily solved obtaining for 
the background gauge fields:
\begin{eqnarray}
\bar{A}(\vecr) &=& -\frac{ig}{4} ( \bar{n} + \frac{\theta}{2\pi}\bar{m}) r^2\\
\bar{A}_D(\vecr) &=& \frac{i}{4g} \bar{m} r^2 ~~,\nonumber
\end{eqnarray}
which reproduces the usual harmonic form for the neutralizing background.

We can now integrate over the fluctuating field $s_{\beta}$, eq.(12),
 by taking into account the constraint given by eq.(7), and obtain:

\begin{eqnarray}
Z [ \bar{n}, \bar{m}; \mu , \nu] = 
\int {\cal D} a
\int {\cal D} a_D e ^{ - S_f [ a, a_D]}
\exp ( i \int d^2r (\nu(\vecr) + \\
\nonumber
+ \frac{\theta}{2\pi} \mu(\vecr)( \bar{A} (\vecr) + 
a(\vecr)) - i \int d^2 r \mu(\vecr) ( \bar{A}_D (\vecr) + a_D(\vecr)) ) 
~~,
\end{eqnarray}
where $\mu(\vecr),\nu(\vecr)$ are the fluctuating charges, 

\beq 
S_f [ a, a_D ] = \frac{g}{2} \int d^2 r [ \partial_\beta ( a_D) ]^2 -
 i\epsilon_{\beta\gamma}\int d^2r [ \partial_\beta (a)
\partial_\gamma (a_D) ]  
\eneq

and we define the background term $S_B$ as:
\begin{eqnarray}
S_B= -i \int d^2r [ \nu(\vecr) + \frac{\theta}{2\pi} \mu(
\vecr) ]  \bar{A} (\vecr) +
 i \int d^2 r \mu(\vecr) \bar{A}_D (\vecr) = \\
\nonumber
-\frac{g}{4} (\bar{n} + 
\frac{\theta}{2\pi}\bar{m}) \int d^2r r^2 (\nu(\vecr) + \frac{\theta}{2\pi} 
\mu(\vecr)) - \frac{1}{4g}\bar{m}\int d^2r r^2\mu(\vecr)~~.
\end{eqnarray}
Eq.(16) gives a ``background term" whose meaning may be derived by 
rewriting the partition function at fixed backgrounds $\bar{N} = \int 
d^2r~
\bar{n}(\vecr)$ and $\bar{M} = \int d^2r~
\bar{m}(\vecr)$ where the fluctuating charges are given by:

\[ 
\mu(\vecr) = \sum_i \mu_i \delta (\vecr - \vecr_i)~,~
\nu(\vecr) = \sum_i \nu_i \delta (\vecr - \vecr_i)~~.
\]

We then get:
\beq
Z [ \bar{N}, \bar{M}, \{\mu\}, \{\nu\}] = 
Z_f e^{-{S_B}} \langle\prod_i exp\left( i \sqrt{g}(\nu_i + 
\frac{\theta}{2\pi}\mu_i)a(\vecr_i) - \frac{i}{\sqrt{g}}\mu_i 
a_D (\vecr_i)\right)\rangle~~,
\eneq
where

\[ 
Z_f \equiv \int {\cal D}a {\cal D}a_D
e^{-S_f [ a, a_D]}
\]

and $\langle~\rangle$ denotes the averaged value with respect to the 
``weight" $\exp(-S_f)$.

In eq.(17) a non neutral correlator between vertices appears. Infact, 
the neutrality condition looks like:

\beq
  \bar{M} = - \sum_{i=1}^{N_p} \mu_i ~~,~ \bar{N} = - \sum_{i=1}^{N_p} \nu_i ~. 
\eneq

In other words the background charges neutralize the Coulomb charges;
therefore the ``splitting" of the charge densities in an uniform 
and a ``fluctuating" part has allowed us to describe a non neutral Coulomb 
gas which generalizes the usual Coulomb gas largely analized in the
 literature \cite{6,9}.

As a result of explicit RG analysis for this 
generalized Coulomb gas (at first order in the relevant parameters) 
we 
find that the IR stable fixed points correspond to the maximum value 
of the critical exponents as it should be.

Being $\bar{M}$ the magnetic background and $\bar{N} $ the electric one 
we assume that the most probable condensate of $N_p$ particles is the one 
which corresponds to the maximum value of the critical exponent which, 
in the presence of magnetic and electric background, is given by:

\beq
x(\nu, \mu) = 2 + g \left[ \nu(\bar{N}-\nu) + \frac{\theta}{2\pi} 
\mu (\bar{M}-\mu)\right] + \frac{1}{g} \mu(\bar{M}-\mu)~~.
\eneq
Then we maximize the above exponent with respect to $\mu$ and 
$\nu$ by imposing the double constraint, given by eq.(18), obtaining:

\begin{eqnarray}
\mu_k = \bar{\mu} = -  \frac{\bar{M}}{N_p}~~~\forall k = 1, \ldots, N_p
\\
\nonumber
\nu_k = \bar{\nu} = - \frac{\bar{N}}{N_p} ~~~\forall k = 1, \ldots, N_p
\end{eqnarray}
We see that the charges of the condensate are the same for each particle
and their values given above are fixed by the 
background only. 

Furthermore it is easy to show that the model is invariant under the 
following discrete $SL(2,Z)$ transformations defined 
 in terms of the complex 
variable $\zeta = 1/g + i\theta/2\pi$ as:

\[
S : \zeta\rightarrow \zeta -i;~~~~ \bar{N}\rightarrow \bar{N} + \bar{M},~~
\bar{M} \rightarrow \bar{M}\hspace{3cm}(21.a)
\]

\[
T : \zeta \rightarrow -\frac{1}{\zeta};~~~~ \bar{N} \rightarrow \bar{M},~~
\bar{M} \rightarrow - \bar{N}
\hspace{3.4cm}(21.b)
\]

\[
C : \zeta \rightarrow \zeta^*;~~~~ \bar{N} \rightarrow - \bar{N},~~
\bar{M} \rightarrow \bar{M}~~.
\hspace{3cm}(21.c)
\]

The above infinite discrete symmetry $SL(2,Z)$ allows us to generate all 
non trivial vacua(IR fixed points) starting from a given one. In fact
the system defined by the background charges $\bar{M}$, $\bar{N}$
transforms in a new one as given by eqs.(21.a,b,c).
At this stage we can compare the previous discrete symmetry $SL(2,Z)$
of our model with the phenomenological "laws of corresponding states"
introduced in ref.[5]:

\setcounter{equation}{21}
1)$f \rightarrow f+1$

2)$ 1/f \rightarrow 1/f +2$

3)$ f \rightarrow 1-f $  ( for$f < 1$)~~,

where the filling factor $f=N_e/N_s$ rewritten in our units is given by
$f=\bar{N}/\bar{M}$.
We then easily proove that the above laws can be expressed in terms of the 
duality transformation, eqs.(21.a,b,c) as 
1)=$S $,
2)=$ TS^2T$,
3)=$ S C$.

To be precise the above transformation laws
generate only a subgroup of the duality transformations, which 
preserves both the oddness of the denominator in the filling $f$ 
and the sign of the condensed electric and magnetic charges(\cite{5}). 
All the previous properties of the model here proposed seem to generate
the complete phase diagram of the Quantum Hall system and its 
hypothized universality properties(\cite{3,5,10}).

Here we only briefly discuss some properties of the critical points 
postponing a detailed analysis to a 
subsequent paper.

Let us start by considering the IR fixed points previously found. We 
have proved that they are stable against CFT 
perturbations and define non trivial field theories, which have the 
following properties:
for a subset of them the charges $Q_e$, $Q_m$, previously defined
satisfy  the ``chirality" condition:
$Q_e = Q_m ~~~~,~\nu = 0$,
i.e. the field theory is chiral and describes the plateaux at filling 
$f=1/\mu$, $\mu$ odd.
Then the relevant vertex operators of the left (chiral) sector
associated to these charges take the simple form
\beq
\hat{V}_\mu (z) = \exp(i\frac{\mu}{\sqrt{g}} \phi_L (z))
\eneq
when expressed in terms of a scalar field $\phi_L$ depending only on 
$z$ (\cite{1}).

Furthermore for these plateaux it is known that
 $\sigma_H$ is a topological invariant
which takes the values
$\sigma_H = \frac{1}{\mu} = \frac{Q_e}{Q_m}$
if one imposes periodic boundary conditions(\cite{11}).

Finally these points are described by a CFT, whose primary 
fields are the
vertices of eq.(22) realized in terms of a chiral field $\phi_L (z)$ 
compactified on a radius $R$ such that $R^2 = \mu$, with central
charge of Virasoro algebra $c$ equal to one.

The simplest IR non trivial fixed point, which corresponds to $f=1$  is 
given by ( $1/\bar{g}=0$, $\bar{\theta}/ 2\pi = -1$). It is not difficult 
to show that the renormalization group flow for the two parameters($g$,
$\frac{\theta}{2\pi}$) defines a circle given by:

\beq
(\frac{1}{g}-\frac{1}{2})^2 +(\frac{\theta}{2\pi}+ 1)^2 = 
\frac{1}{4}~~,
\eneq

which has radius equal to $1/2$ and is tangent to the $1/g$ axis at 
$\bar{\theta}/2\pi=-1$. The interior of the circle defines a phase 
whose attractive point is the above one and where the condensate is a
dyon with electric and magnetic charge equal to unity and $\sigma_H =1$.
Then by using the duality transformations given by eqs.(21.a,b,c)
we can get all the other non trivial
IR fixed points associated to the fillings $\nu = p/\mu$ (with $p$ and $\mu$ prime
factors). 
The $SL(2, Z)$ matrix which realizes that is determined by the finite
fraction representation of the rational $p/\mu$ (\cite{7})

The same matrix maps circles into circles in such a way that the tangent 
points of two of them are unstable repulsive fixed points describing the
physics of the transition region between plateaux.  

Then all these fixed points are in the same universality class of the $c=1$ CFT
just mentioned; 
that is a crucial point concerning the universal character of the 
transition  between two plateaux. 
In particular it would be very interesting to find the appropriate 
operator which, in Conformal Field Theory, drives such a transition. 
Then by the use of the well known techniques of 2D CFT we should be 
able to evaluate explicitly the critical exponents.

Finally, a clever use of the duality transformation laws (21a,b,c) 
should give us precise informations about the other phases (as for 
ex. the Hall insulator one) of our physical system.

\vspace{2.5cm}

{\bf Acknoledgement}

We thank R. Musto and F. Nicodemi for the many discussions we have had 
about the Quantum Hall Effect during the long period of collaboration with
two of us [G. C. and G. M.]. We wish also to thank L. Peliti for his interest
in our work and for a critical reading of the manuscript.

\end{document}